\documentclass[conference]{IEEEtran}
\IEEEoverridecommandlockouts
\usepackage{cite}
\usepackage{amsmath,amssymb}

\DeclareMathOperator*{\argmin}{arg\,min}
\usepackage{algorithmic}
\usepackage{graphicx}
\usepackage{caption}
\usepackage{subcaption}
\usepackage{textcomp}
\usepackage{xcolor}
\usepackage{nicefrac}
\usepackage{bm}
\usepackage{hyperref}
\usepackage{tabularx}
\usepackage{booktabs}

\def\BibTeX{{\rm B\kern-.05em{\sc i\kern-.025em b}\kern-.08em
    T\kern-.1667em\lower.7ex\hbox{E}\kern-.125emX}}

\begin{document}

\onecolumn 
{\Huge \bfseries  IEEE Copyright Notice}

\vspace{20mm}

\large This work has been submitted to the IEEE for possible publication. Copyright may be transferred without notice, after which this version may no longer be accessible.

\newpage

\twocolumn

\title{Surf-CDM: Score-Based Surface Cold-Diffusion Model For Medical Image Segmentation}

\author{Fahim Ahmed Zaman, Mathews Jacob, Amanda Chang, Kan Liu, Milan Sonka and Xiaodong Wu
\thanks{Fahim Ahmed Zaman, Mathews Jacob, Milan Sonka and Xiaodong Wu are with the Department of Electrical and Computer Engineering, University of Iowa, Iowa City, IA 52242, USA (e-mails: \{fahim-zaman, mathews-jacob, milan-sonka, xiaodong-wu\}@uiowa.edu)}
\thanks{Amanda Chang is with the Division of Cardiology, Department of Internal Medicine, University of Iowa, Iowa City, IA 52242, USA (e-mails: amanda-chang@uiowa.edu)}
\thanks{Kan Liu is with the Division of Cardiology and Heart and Vascular Center, School of Medicine, Washington University in St Louis, St. Louis, MO 63130, USA (e-mails: kanl@wustl.edu)}
}

\maketitle

\begin{abstract}
Diffusion models have shown impressive performance for image generation, often times outperforming other generative models. Since their introduction, researchers have extended the powerful noise-to-image denoising pipeline to discriminative tasks, including image segmentation. In this work we propose a conditional score-based generative modeling framework for medical image segmentation which relies on a parametric surface representation for the segmentation masks. The surface re-parameterization allows the direct application of standard diffusion theory, as opposed to when the mask is represented as a binary mask. Moreover, we adapted an extended variant of the diffusion technique known as the ``cold-diffusion'' where the diffusion model can be constructed with deterministic perturbations instead of Gaussian noise, which facilitates significantly faster convergence in the reverse diffusion. We evaluated our method on the segmentation of the left ventricle from 65 transthoracic echocardiogram videos (2230 echo image frames) and compared its performance to the most popular and widely used image segmentation models. Our proposed model not only outperformed the compared methods in terms of segmentation accuracy, but also showed potential in estimating segmentation uncertainties for further downstream analyses due to its inherent generative nature.
\end{abstract}


\section{Introduction}
In the era of precision medicine, image segmentation is critical for image-derived quantitative measures that support clinical decisions in diagnosis, treatment, surgical planning, image guided surgery etc. \cite{Preeti}. The advances in deep learning (DL) based automated image segmentation have introduced various highly efficient methods that achieved dominance over traditional edge detection and mathematical models \cite{Hesamian, wang2022}. These methods mostly include convolutional neural networks (CNN), vision transformers (ViT) and graph-based neural networks, which are generally trained end-to-end in a discriminative manner. On the other hand, conditional generative models have also emerged for image segmentation which have the advantage of learning the underlying statistics of segmentation masks. These conditional generative models can be classified into two major categories: generative adversarial networks (GAN) and the diffusion probabilistic models (DPM). 

The DPM have shown remarkable results for image generation in computer vision applications, often times outperforming GAN and other generative models \cite{dhariwal2021diffusion}. Recently, a lot of research has been driven towards adapting the diffusion models for medical image segmentation \cite{wu2022medsegdiff,wu2023medsegdiff,wolleb2021diffusion,Rahman2023AmbiguousMI}. These methods are mainly based on score-based diffusion models \cite{song2021scorebased} which consist of three major elements: (1) a forward diffusion process, (2) score-based denoiser and (3) a reverse diffusion process. The objective of the forward diffusion process for image segmentation differs from that of the image generation, such that, the goal is to perturb the segmentation mask with Gaussian random noise given a noise scale, while the image remains unchanged. The denoiser is trained to learn the scores for noisy data distribution of all the noise scales. The goal of the reverse diffusion process is to sample segmentation masks from a distribution of given noise scale, conditioned on the image. 

Injecting Gaussian noise directly to the segmentation masks poses a notable disadvantage due to the unnatural distortion it introduces to the underlying distribution. It can be argued that learning the statistics of segmentation masks, characterized by being bimodal or having very few modes depending on the number of semantic classes, is challenging because there is no smooth transition between class modes. One remedy for this issue is to re-parameterize the discrete representation of the segmentation masks to a continuous one. Bogensperger {\em et al.} proposed to transform the discrete segmentation mask to signed distance function (SDF), where each pixel represents the signed euclidean distance from the closest object boundary \cite{bogensperger2023scorebased}. Then a conditional score-based diffusion model is developed using the SDF representations. In the reverse diffusion process, the final segmentation mask is obtained by distance thresholding. Bansal {\em et al.} \cite{bansal2022cold} demonstrated that diffusion models  can be extended beyond the theoretical frameworks of the standard setting. In these extended cold-diffusion models, there is no requirement for the forward and reverse diffusion process to be specifically designed around Gaussian noise. The perturbations can be either randomized or deterministic. 

Inspired by the cold-diffusion models, we propose a novel score-based conditional cold-diffusion method for medical image segmentation, that incorporates graph-based parametric surface representation technique as its initial step. The contributions of this work can be summarized, as follows.
\begin{enumerate}
  \item To the best of our knowledge, this is the first work to leverage the cold-diffusion model for medical image segmentation. As the noise space is constrained by the deterministic perturbations, fewer sampling steps are needed than the standard diffusion models to produce accurate segmentation,
  \item We parameterized the discrete segmentation mask to a graph structure, often used in graph-based surface segmentation methods \cite{chen2018, xie2022globally, wu2023model}. The parametric surface representation allows direct incorporation of standard diffusion techniques for forward and reverse diffusion,
  \item Due to the generative nature of the method, segmentation uncertainty can be estimated for further downstream analyses (i.e., fine-tuning using interactive image segmentation \cite{ramadan2020survey}).
\end{enumerate}

\section{Proposed Method}
We propose a score-based surface cold-diffusion model (Surf-CDM) having four major steps, (1) Graph-based parametric surface representation: We construct a graph structure from a given segmentation mask, maintaining pixel neighborhood relations, (2) Forward Diffusion: We use a deterministic degradation operator in the forward process for segmentation mask perturbation in the constructed graph, (3) Score-based denoiser: We learn a score-based model for score matching of the perturbations using an U-net shaped denoiser, (4) Reverse diffusion: We use Langevin dynamics to draw samples using the learned score-based model.

\subsection{Graph-based parametric surface representation}

Let $I_t$ and $m$ of size $X \times Y$ be the image and 2-D binary mask, respectively. For each $x$ (i.e., $x \in X$), the pixel subset {$m(x,y) \mid 0 < y \leq Y$} forms a column parallel to y-axis, denoted by $\mathcal{C}(x)$. Each column has a set of neighboring columns for a certain neighboring setting $N$, e.g., the four-neighbor relationship. Let the boundary surface $S$ of the target object mask $m$, is a terrain-like surface and intersects each column $\mathcal{C}(x)$ at exactly one pixel. Thus $S$ can be defined as a function $S(m)$, which maps each column $\mathcal{C}(x)$ to the surface position (y-value) on the column. 
\begin{figure}[h]
    \centering
    \includegraphics[width=3.3in]{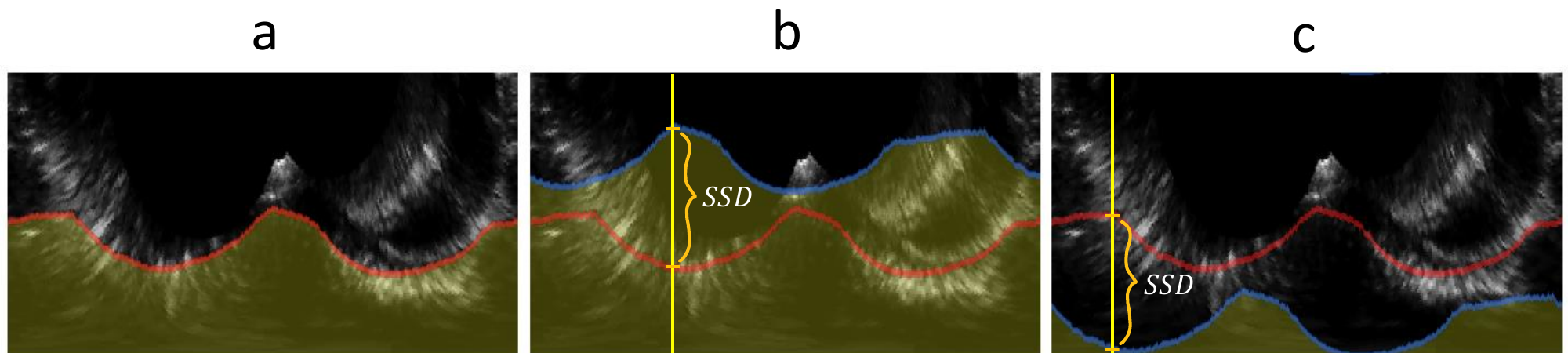}
    \caption{\textbf{a)} An example image $I_t$ with segmentation mask $m$ (yellow), where red line is the reference surface $S$. Each column of $m$ intersects with $S$ only once. \textbf{b)-c)} Two example of mask perturbation for a $\sigma_{i}$ ($1\leq i \leq n $), where b) shows an over-segmentation and c) shows an under-segmentation. In each sub-figure, the yellow region shows the object area, the blue line shows the perturbed surface $\tilde{S}$ and the light yellow line shows an example column, where SSD indicates the surface-to-surface distance on the column.}
    \label{Figure1}
\end{figure}
The advantage of working with $S$ over $m$ is that for any deterministic perturbation, the surface-to-surface distance of the reference surface $S$ and the perturbed surface $\tilde S$ is a real number for each column, where $m$ is a discrete representation and often times bimodal (having only one object class with background). Moreover, with surface parameterization, we can adapt the sampling process of the standard diffusion models maintaining surface monotonicity. An example of surface parameterization is shown in Fig. \ref{Figure1}-a.

\begin{figure*}[ht!]
    \centering
    \includegraphics[width=6in]{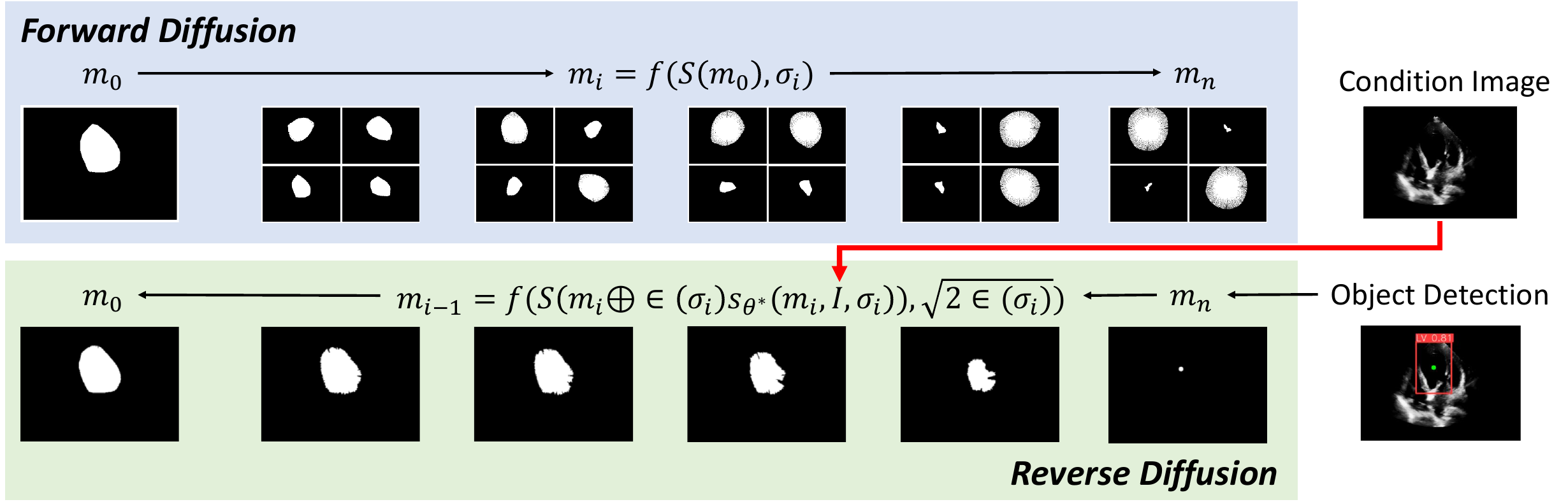}
    \caption{The blue block demonstrates forward diffusion process. Top row (left to right) shows the distortion for $i=0,1,\dots n$ on a segmentation mask $m$ of an image $I$. $4$ randomly sampled perturbation for each step is shown. Note that, the natural distortion results in either over-segmentation or under-segmentation in each step depending on the noise scale $\sigma$. The green block demonstrates reverse diffusion process (bottom row right to left) which starts with object detection and then uses Langevin MCMC iterative process conditioned on the image $I$. Although the denoiser takes parametric masks and images as input, the output masks and images are shown in the original image domain for visual convenience.}
    \label{Figure2}
\end{figure*}

\subsection{Forward Diffusion}
Let $m$ denotes a segmentation mask. Consider a sequence of noise scales $\sigma_1 < \sigma_2 < \dots < \sigma_n$, where $\sigma_1$ is small enough such that $p_{\sigma_{1}} (m) \approx p_{data}(m)$. In the standard diffusion model, scaled Gaussian noise is injected on the image given $\sigma$. To create natural distortions of $m$, instead, we leverage the cold-diffusion techniques \cite{bansal2022cold} and modeled our degradation operator that generates deterministic perturbations, with 

\begin{equation}
    \label{deterministic_forward_operator_equation}
    \begin{aligned}
        m_i = f(S(m),\sigma_{i}),
    \end{aligned}
\end{equation}
where $f(S(m),\sigma_{i}$) performs vertical shifting and horizontal rotation of surface $S$ for given $\sigma_{i}$ and produce perturbed segmentation $m_{i}$ such that, for $\sigma_1$, the unsigned surface-to-surface distance (USSD) of the surface $S$ and perturbed surface $S_{i}$ is minimum (i.e., USSD $\approx 0$), and for $\sigma_n$, USSD will be very high. Note that, the degradation operator is independent of the image to be segmented, thus it remains unchanged. Fig. \ref{Figure1}-b)-c) shows two examples of forward diffusion. 

\subsection{Score-based denoiser}
For image generation, Song {\em et al.} \cite{song2019generative} proposed to train a Noise Conditional Score Network (NCSN), denoted by $s_\theta(\mathcal{I},\sigma)$ with a weighted sum of denoising score matching \cite{vincent2011connection}, where the score function of a distribution $p_{\sigma}(\mathcal{I})$ is defined as $\nabla_{\mathcal{I}}\log p_{\sigma}(\mathcal{I})$. For the segmentation task, we can denote a conditional NCSN as $s_\theta(m,I_t,\sigma)$. Using Tweedie's formula \cite{efron2011tweedie}, we minimize the following objective, 

\begin{equation}
    \label{segmentation_objective}
    \begin{split}
    \theta^* = \argmin_{\theta} & (\sum_{i=1}^{n}\lambda(\sigma_{i}) \mathbb{E}_{p_{data_{m}}(m_{i} \mid m)}\mathbb{E}_{p_{\sigma_{i}}(m_{i} \mid m)} \\& \times [\Vert s_{\theta}(m_{i},I_t,\sigma_{i})-\nabla_{m_{i}}\log p_{\sigma_{i}}(m_{i}\mid m) \Vert^{2}_{2}]),
    \end{split}
\end{equation}
where $\lambda(\sigma_{i}) \propto 1/\mathbb{E}[\Vert \nabla_{m_{i}}\log p_{\sigma_{i}}(m_{i} \mid m) \Vert]$ is a positive weighting function, $I_{t}$ is the image to be segmented.

\begin{figure}[h]
    \centering
    \includegraphics[width=3in]{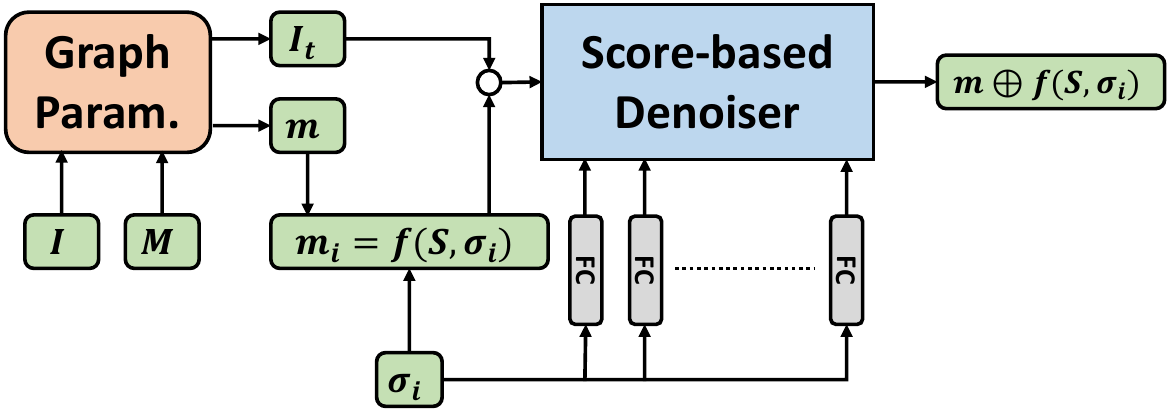}
    \caption{Denoiser training framework. $I$ \& $M$ are the image and segmentation mask. The grarph param block re-parameterize $I$ \& $M$ to $I_{t}$ \& $m$, respectively. $\sigma_{i}$ is the noise modulator. $f(S,\sigma_{i})$ is the perturbed mask given surface $S$ \& $\sigma_{i}$. Noise modulator is connected to each convolution kernel of the denoiser using fully connected (FC) blocks. The denoiser outputs the added perturbation. Here, $\oplus$ is an elementwise xor operator.}
    \label{Figure3}
\end{figure}

An auto-encoder is trained to learn the score-function $\nabla_{\tilde m}\log p_{\sigma_{i}}(\tilde m\mid m)$. A conditioning on the image $I_t$ is required, for which we concatenated the conditioning image $I_t$ with the corresponding segmentation $\tilde{m}$, roughly following recent works related to conditional generative modeling \cite{ho2022cascaded,saharia2022palette,zaman2023trust,JMI}, and used as a 2-channel input to the network. The other input to the network is the noise modulator $\sigma$. An overview of the denoiser training is shown in Fig. \ref{Figure3}.

\subsubsection{Denoiser architecture}
\label{architecture}
The denoiser is a U-net \cite{ronneberger2015u} shaped auto-encoder which is adapted from \cite{CIBM}. The noise modulator $\sigma$ is connected to fully connected layers and then multiplied with each hidden convolution layers. Here, the number of fully connected layers are same as the number of hidden convolution layers. The nodes of each of the fully connected layers are same as the number of kernels of the convolution layer they are connected to.

\subsection{Reverse diffusion}
Once the score-based model is trained given sufficient data and model capacity, the optimal $s_\theta(m,I_t,\sigma)$ matches $\nabla_{\tilde m}\log p_{\sigma_{i}}(\tilde m\mid m)$ for all $\sigma \in \{\sigma_i\}_{i=1 \dots n}$. Song {\em et al.} \cite{song2019generative} proposed to run iterative steps of Langevin Markov chain Monte Carlo (MCMC) to get a sample for each $p_{\sigma_{i}}(m)$ sequentially. We formulate the iterative sampling as,

\begin{equation}
    \label{langevin_dynamic_equation}
    m_{i-1} = f(S(m_{i} \oplus (\epsilon(\sigma_{i})s_{\theta^{*}}(m_{i},I_t,\sigma_{i}))),\sqrt{2\epsilon(\sigma_{i})}),
\end{equation}
where $i=n, n-1, \cdots, 1$, $\epsilon \propto \sigma$ is a positive weighted scaler that controls the step size. As $i \rightarrow 1$ \& $\epsilon(\sigma_i) \rightarrow 0$, $m_{0}$ becomes a sample from $p_{\sigma_{i}}(m) \approx p_{data}(m)$. 

\section{Experiments}

\subsection{Dataset}
To demonstrate the effectiveness of our proposed Surf-CDM method, echocardiogram video (echo) dataset from The University of Iowa hospitals is selected for performance evaluation. The echos are acquired by Transthoracic echocardiography (TTE) using standard techniques of 2D echocardiography following the guidelines of the American Society of Echocardiography \cite{lang2015recommendations}. All the echos have the standard apical 4-chamber left ventricular (LV) focused view. In total, the selected dataset contains 65 echos (30 of Takotsubo patients, 35 of ST Elevation Myocardial Infarction patients), with 2230 still-frame gray-scale images ($18 \sim 112$ frames per video). The LVs were manually traced fully by an expert using ITK-snap.

\begin{figure}[h]
    \centering
    \includegraphics[width=3in]{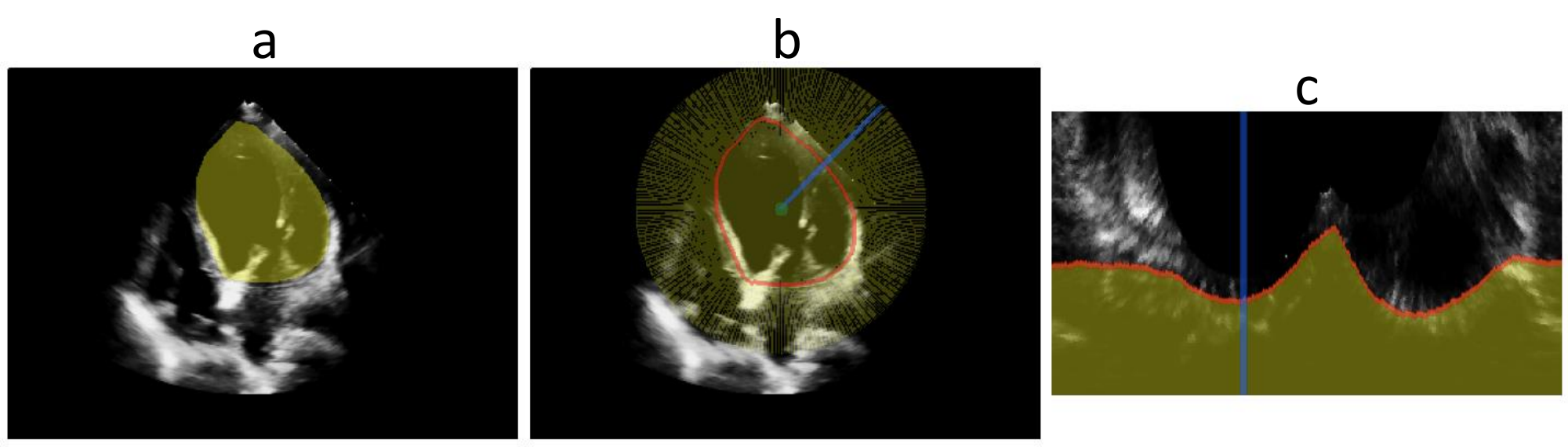}
    \caption{\textbf{a)} An example image $\mathcal{I}$ with Left-Ventricle (LV) segmentation mask $\mathcal{M}$ (yellow) overlay. \textbf{b)} Polar co-ordinate transformation for $\mathcal{I}$ \& $\mathcal{M}$. Yellow lines shows the graph columns with red dot as the centroid. \textbf{c)} Transformed image $I_{t}$ \& mask $m$ (yellow). Each column of $I_{t}$ \& $m$ represents graph column which intersects with surface $S$ (red) only once. Blue line in b) and c) shows the same graph column before and after sampling, respectively.}
    \label{Figure4}
\end{figure}

\subsubsection{Pre-processing}
\label{preprocessing}
The images were originally stored at a resolution of $800 \times 600$ pixels and later resized to $256 \times 352$ for training and inference. Due to the intensity variations in the data, augmentation was performed using intensity variation ($-0.1 \sim 0.1$ of the maximum image intensity) along with random horizontal flips. Let $\mathcal{I}$ and $\mathcal{M}$ be a given image and a 2-D binary segmentation mask of the dataset with a size $A \times B$, respectively. $c(a,b)$ is the centroid of the object region, where $0<a \leq A$ and $0<b \leq B$. For surface parameterization, we use polar co-ordinate transformation taking $c(a,b)$ as the center of the cartesian plane. The column length is set to $200$ pixels. In the training phase, reference segmentation masks are used for surface parameterization. In the inference phase, $c(a,b)$ is unavailable. So, an object detector (\ref{sampling_process}) is trained to generate the bounding box of the object with a center $c(a,b)$, which can be used for surface parameterization.

\subsection{Training parameters}
For the diffusion parameters, we set $\sigma_{1}=0.1$, $\sigma_{n}=1.0$ and number of steps $n=10$. We used mean-squared-error as the loss function and Adam as the optimizer with a learning rate of $1e^{-3}$ to train the denoiser. The dataset was split into $70:10:20$ ratio for training, validation and testing maintaining no overlapping of the same patients in different categories.

\subsection{Sampling in the reverse diffusion}
\label{sampling_process}

To initialize the segmentation process, we need an approximate centroid inside the target object. We trained YOLOv5+ \cite{glenn_jocher_2022_7347926} as an object detector, which is a popular object detection model in computer vision, to generate a bounding box around the LV. The center of the bounding box is used as the approximation of the LV centroid for surface parameterization. 

To obtain segmentation masks $\tilde{m}$ for test images $I_{t}$, we used Eq. (\ref{langevin_dynamic_equation}). Using $n=10$ was enough to produce accurate segmentations. In our experiments, increasing the value of $n$ did not increased the accuracy.

\section{Results}

\begin{table}[ht!]
    \footnotesize
    \caption{Quantitative results for LV segmentation of the echo dataset. }
    \label{table-I}
    \resizebox{\columnwidth}{!}{
    \begin{tabular}{llll}
        \toprule
		\textbf{Method} & \textbf{DSC $\uparrow$} & \textbf{IoU} $\uparrow$& 95\%\,\textbf{HD} $\downarrow$ \\
  		\textbf{} & \textbf{($Mean \pm SD$)} & \textbf{($Mean \pm SD$)} & \textbf{($Mean \pm SD$)}\\
        \midrule
		U-net \cite{ronneberger2015u} & $0.863 \pm 0.090$ & $0.770 \pm 0.133$ & $3.718 \pm 0.736$ \\
        V-net \cite{milletari2016v} & $0.927 \pm 0.036$ & $0.865 \pm 0.059$ & $3.338 \pm 0.630$ \\
        Res-Unet \cite{xiao2018weighted} & $0.931 \pm 0.027$ & $0.873 \pm 0.046$ & $3.277 \pm 0.647$ \\
        DeepLabV3+ \cite{chen2018encoder} & $0.932 \pm 0.028$ & $0.874 \pm 0.047$ & $3.335 \pm 0.534$ \\
        \textbf{Surf-CDM (Ours)} & $\mathbf{0.940 \pm 0.019}$ & $\mathbf{0.887 \pm 0.033}$ & $\mathbf{3.224 \pm 0.579}$ \\
        \bottomrule
    \end{tabular}}
\end{table}

We used 3 standard metrics to evaluate the segmentation results: (1) Dice similarity co-efficient score (DSC), (2) Intersection over union (IoU) and (3) $95\%$ Hausdorff distance (HD). Table 1 shows the quantitative results for the echo dataset segmentation with different methods. The results indicate that our proposed method outperforms all the compared methods for all three metrics.

\section{Discussion}

\begin{figure}[h]
    \centering
    \includegraphics[width=3.3in]{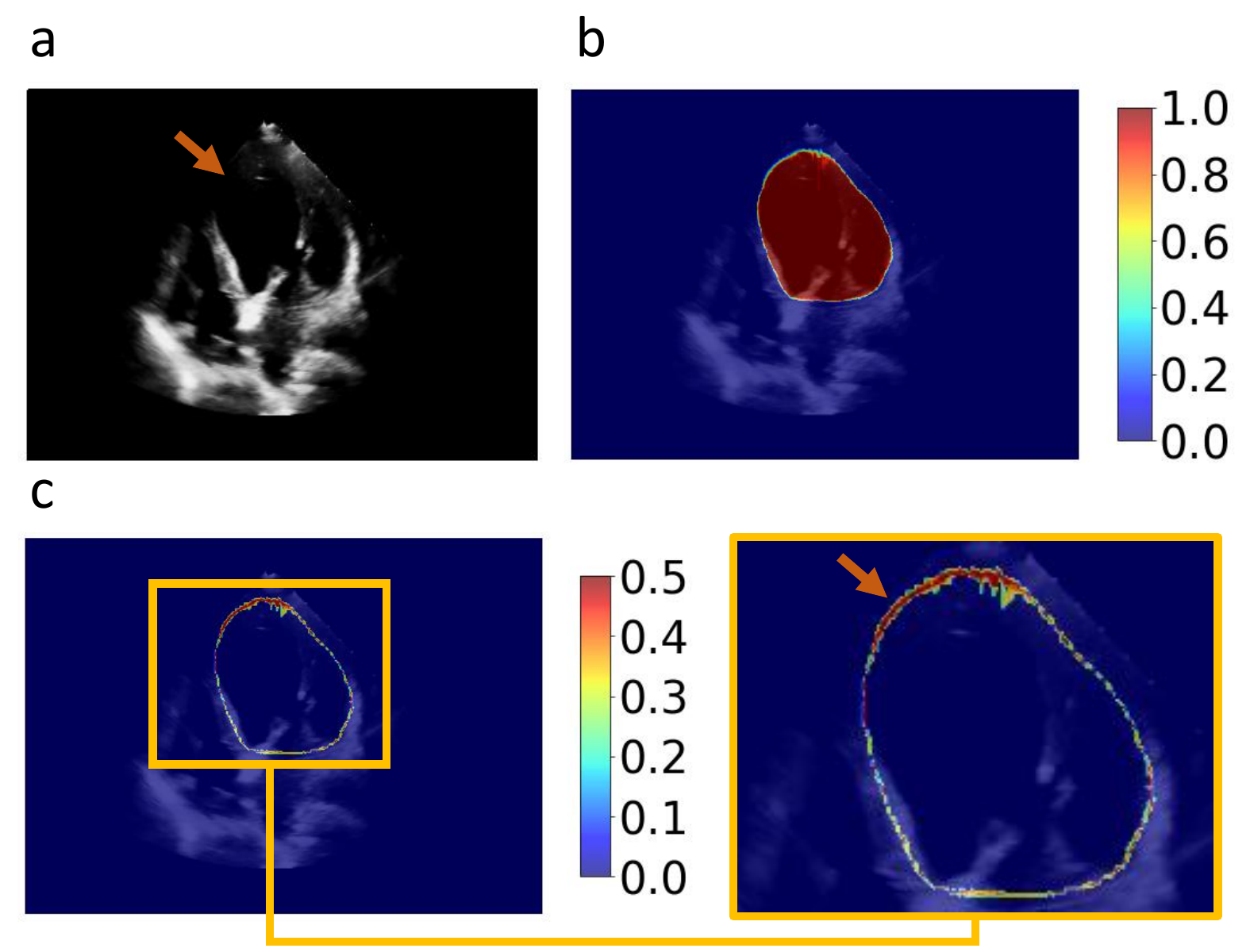}
    \caption{\textbf{a)}. A sample image with orange arrow showing the missing boundary of LV. \textbf{b}. Mean of the predicted segmentations of 20 sampling runs overlaid on the image. \textbf{c)}. SD (uncertainty) of the predicted segmentations of 20 sampling runs. The orange arrow showing the most uncertain area. Notice that the orange arrows in a) and c) corresponds to the same location where the LV boundary is missing.}
    \label{Figure5}
\end{figure}

Due to the generative nature of Surf-CDM, we can estimate segmentation uncertainty by calculating standard deviation (SD) over a fixed sampling runs. Fig. \ref{Figure5} shows an example of segmentation uncertainty estimation of a sample image. Notice that the high uncertainty in the region without a clear  object boundary. This information may be used to  further fine-tune the segmentation boundary by an human expert using interactive segmentation methods. Another thing to note is that Surf-CDM needs only 10 iterations in the reverse diffusion to produce accurate segmentations, whereas the standard diffusion models usually take more than 100 iterations. This indicates leveraging cold-diffusion for deterministic perturbation on re-parameterized surface yields much faster convergence than directly injecting Gaussian noise on discrete segmentation mask.

Any star-shaped object can be easily re-parameterized for graph construction using polar-coordinate transformation. For more complex structured objects, advanced unfolding techniques \cite{wu2023model,DEKRAKER2018408} can be applied for surface parameterization. A limitation of this method is that it may be nontrivial to be extended for multiple object segmentation. One probable solution to this problem is to add different channels in the denoiser output for simultaneous prediction of multiple object classes or layers (i.e., OCT layer segmentation) while maintaining mutual interaction between objects, which is left for our future work. 

\section{Conclusion}
The general DL-based segmentation models may frequently fail in medical image segmentation due to various uncertainty, such as, complex tissue structures, noisy acquisition, disease-related pathologies, as well as the lack of sufficiently large datasets with associated annotations. Our proposed Surf-CDM method not only outperformed the standard methods in terms of segmentation accuracy, but can also capture the underlying uncertainty for further investigation and fine-tuning.

\section{Acknowledgments}
This research was supported in part by NIH the NIBIB Grants R01-EB004640, R01-AG067078 and R01-EB019961. There is no other conflicts of interest.

\section{Compliance with ethical standards}
This study was performed in line with the principles of the
Declaration of Helsinki and approved by the local institutional review board.

\bibliographystyle{IEEEtran}
\bibliography{ISBI2024}

\end{document}